\documentclass[preprintnumbers,showpacs,twocolumn,amsmath,amssymb]{revtex4}
\usepackage{color}
\usepackage{times}
\usepackage{epsf}
\usepackage{graphicx}
\usepackage{dcolumn}
\usepackage{bm}

\begin{document}
\draft

\title{Appearance of Mobility Edge in Self-Dual Quasiperiodic Lattices}

\author{Gang Wang$^{1}$, Nianbei Li$^{2}$, and Tsuneyoshi Nakayama$^{1,3}$}
\affiliation{$^{1}$Max Planck Institute for the Physics of Complex
Systems, N\"{o}thnitzer Stra\ss e 38, 01187 Dresden, Germany}
\affiliation{$^{2}$Center for Phononics and Thermal Energy Science,
School of Physics Science and Engineering, Tongji University, 200092
Shanghai, China} \affiliation{$^{3}$Hokkaido University, 064-0828
Sapporo, Japan}


\begin{abstract}

Within the framework of the Aubry-Andr\'{e} model, one kind of
self-dual quasiperiodic lattice, it is known that a sharp transition
occurs from \emph{all} eigenstates being extended to \emph{all}
being localized. The common perception for this type of
quasiperiodic lattice is that the self-duality excludes the
appearance of the mobility edge separating localized from extended
states. In this work, we propose a multi-chromatic quasiperiodic
lattice model retaining the self-duality identical to the
Aubry-Andr\'{e} model, and demonstrate numerically the occurrence of
the localization-delocalization transition with definite mobility
edges. This contrasts with the Aubry-Andr\'{e} model. As a result,
the diffusion of wave packet exhibits a transition from ballistic to
diffusive motion, and back to ballistic motion. We point out that
experimental realizations of the predicted transition can be
accessed with light waves in photonic lattices and matter waves in
optical lattices.

\end{abstract}


\maketitle

\newpage


The concept of Anderson localization-delocalization (LD) transition
has been progressively developed from its original scope of
solid-state physics to a broad class of physical systems, including
light waves in photonic lattices~\cite{ALLight,LahiniPRL2008},
matter waves in optical
potentials~\cite{BillyNature2008,RoatiNature2008}, sound waves in
elastic media~\cite{HuNP2008}, and quantum chaotic
systems~\cite{GarreauPRL2008}.
It is a quantum phase transition caused by disorder where waves
experience from  delocalized (``metallic" phase) to exponentially
localized (``insulator" phase)
states~\cite{AndersonPR1958,GangofFour,LeeRMP1985,Kramer1993}. This
concept predicts a wealth of interesting phenomena. In usual
one-dimensional (1D) disordered systems, all eigenfunctions are
localized, so there is no LD transition~\cite{Notes2}. While in
three-dimensional (3D) disordered systems there should exist a
transition between delocalized and localized states at a
well-defined critical energy called the mobility edge
$E_c$~\cite{GangofFour}.


The absence of the LD transition in usual 1D disordered systems
makes itself trivial from the perspective of the physics of
disorder-induced localization transition. However, Aubry and
Andr\'{e} ~\cite{AA} have proposed self-dual quasiperiodic (QP)
lattices, the so-called Aubry-Andr\'{e} (AA) model (also known as
the Harper model~\cite{Harper}), showing localization transition
where the QP potential of finite strength mimics the deterministic
disorder. One of the key features of the AA model is either
\emph{all} states being extended or localized depending on the
modulation strength of the potential~\cite{SokoloffPR1985}. This
localization transition in the modulation strength space arises from
the self-duality of the AA model. As a result, either ballistic or
localized excitations are observed in the AA
model~\cite{HiramotoJPSJ1988}. Involved physics in the AA model has
been extensively investigated, \textit{e.g.}, metal-insulator
transition~\cite{FishmanPRL1982,KohmotoPRL1983,
SarmaPRL1988,SilberbergPRL2009,SarmaPRL2010}, Hofstadter's
butterfly~\cite{Hofstadter}, and topological phase
transitions~\cite{KrausPRL2012}, just to mention a few.
Experimentally, the LD transition of the AA model has been studied
in the context of matter waves in bichromatic optical
lattices~\cite{RoatiNature2008} and light waves in fabricated
photonic lattices~\cite{SilberbergPRL2009}. Both approaches have
clearly demonstrated the existence of the localization transition.

The localization transition in the AA model is unique because all
states are either localized or extended without energy-dependent
mobility edges. However, little attention has been paid to whether
or not the mobility edge can exist in QP lattices, especially under
the situation of self-duality being retained~\cite{Notes1}. Since it
has been generally believed that the self-duality keeps all the
states being localized or extended, one would expect that the
delocalized and localized states do not coexist at the same
modulation strength.

In this work, we demonstrate that the above view is not generally
true by proposing a new type of self-dual QP model. In this model,
the LD transition becomes anomalous, and definite mobility edges can
emerge despite the restriction of self-duality. We shall consider
the photonic lattices, artificially fabricated arrays of
evanescently coupled waveguides, to investigate the transition. The
great advantage of photonic lattices lies in the easy fabrication of
complex refractive-index landscapes, and in the direct observation
of wave function itself during the
transport~\cite{LedererNature2003}.

\begin{figure*}[htb]
\includegraphics[width=1\linewidth]{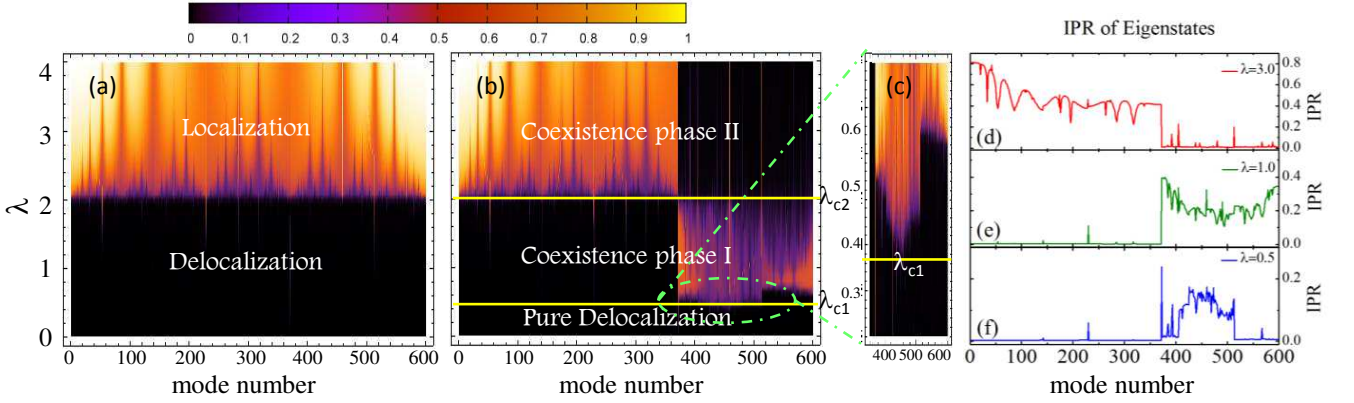}
\caption{(Color online) (a, b): Amplitude profiles of the
eigenstates' IPR of lattices versus the strength of modulation
$\lambda$ (vertical axis). Panel (a) corresponds to the original AA
lattice with $\alpha=(\sqrt{5}-1)/2$ (inverse of golden mean), while
panel (b) to the generalized model Eq.~(\ref{eq3}) with
$\alpha=(\sqrt{5}-1)/2$ and $\sigma=c_2=1/3$. We denote more
extended states by darker shading while more localized states by
lighter shading (see the color bar). (c): Close-up view of the
circled area. (d, e, f): Plots of IPR of individual states for
different $\lambda$'s in Eq.~(\ref{eq3}), namely, $\lambda=0.5, 1.0,
3.0$, to show the mobility edges and anomalous LD transition. All
the above calculations are done for a system with $N=600$ sites, and
the mode number is arranged in descending order.}\label{fig1}
\end{figure*}

The AA model is given by the following eigenvalue equation, $ E
a_n=\lambda \cos (2 \pi \alpha n) a_n + c (a_{n+1}+ a_{n-1}),$ where
$a_n$ is the amplitude of wave function at the lattice site $n$, and
$\lambda$ and $c$ denote the strength of the deterministic disorder
and the site-to-site hopping energy, respectively. The irrational
number $\alpha$ indicates the ratio between the period of the
modulation and the underlying periodic lattice. These types of QP
lattices possess unique character being intermediate between full
periodicity and full disorder. One premier feature of the AA model
is its self-duality, which means that the Fourier transformation of
the above equation, $ E f_k=2c \cos (2 \pi \alpha k)
f_k+\frac{\lambda}{2} (f_{k+1}+ f_{k-1}),\label{eq2} $ definitely
takes the same form as the original one with the roles of $c$ and
$\lambda/2$ being interchanged, transforming localized states into
extended states and vice versa. As a result, there exists a critical
strength $\lambda/c=2$ identified as the transition point to
separate the localization and delocalization phases. Therefore, the
AA model exhibits a transition in parameter space: for $\lambda <
2c$ all the states are extended, while the situation $\lambda > 2c$
makes all of the states exponentially localized. However, this type
of self-dual QP system has no emergent energy-dependent mobility
edges.

To characterize the localization natures in self-dual QP lattices,
we calculate the inverse participation ratio (IPR) index defined by
$P^{-1}=\sum _{i=1}^N \left|a_i\right|^4/\left(\sum _{i=1}^N
\left|a_i\right|^2\right)^2,$ where $N$ being the number of lattice
sites. The IPR represents a measure of the number of sites
contributing to a given state~\cite{Kramer1993}. For spatially
extended states, $P^{-1}$ approaches $1/N$ whereas it is finite for
localized states. Therefore, the IPR can be taken as a criterion to
distinguish the extended states from the localized
ones~\cite{Kramer1993}. Figure~\ref{fig1} (a) shows the IPR values
of all eigenstates for the AA model as a function of $\lambda$. The
clear distinction of $P^{-1}$ indicates the characteristic features
of the AA model. A sharp LD transition occurs at the dual point
$\lambda=2$ (here we set $c$ as energy unit), beyond which all
eigenstates convert from extended to localized. For a fixed
modulation, the (de)localization occurs independently of energies of
the modes. Furthermore, the localization transition does not depend
on the incommensurate ratio $\alpha$. Accordingly, the asymptotic
dynamics of a wave packet will exhibit a ballistic motion for
$\lambda<2$, but will come to a halt for $\lambda>2$. Whereas at
$\lambda=2$ (transition point), the dynamics becomes normal
diffusion~\cite{HiramotoJPSJ1988}.

Now we propose another class of self-dual models with multiple QP
modulations showing the anomalous LD transition and definite
mobility edges. This model incorporates the non-nearest-neighbor
hopping into the multi-chromatic QP
lattices~\cite{SoukoulisPRL1982}. In photonic lattices, the
inclusion of higher-order hopping can be realized and controlled by
arranging the topology of the
arrays~\cite{DreisowOL2008,WangOL2010}. Without loss of generality,
we only retain up to the next-nearest-neighbor hopping $c_2$ besides
the nearest-neighbor hopping $c_1$. This yields
\begin{eqnarray}
E a_n&=&\lambda [\cos (2 \pi \alpha n)+\sigma \cos (4\pi \alpha
n)] a_n\nonumber \\
& &+ c_1 (a_{n+1}+ a_{n-1}) + c_2 (a_{n+2}+
a_{n-2}).\label{eq3}
\end{eqnarray}
The Fourier transformation of Eq.\,(\ref{eq3}) provides the
following equation:
\begin{eqnarray}
E f_k&=&[2 c_1 \cos (2 \pi \alpha k)+ 2c_2 \cos (4 \pi \alpha k)
]f_k\nonumber \\
& &+\frac{\lambda}{2} (f_{k+1}+ f_{k-1})+\frac{\lambda \sigma}{2}
(f_{k+2}+ f_{k-2}).\label{eq4}
\end{eqnarray}
Obviously, the model manifests self-duality under the condition of
$\sigma=c_2/c_1$. From the experience of the AA model, one would
expect that such a self-dual model undergoes a LD transition at
$\lambda=2 c_1$, and the extended and localized states do not
coexist. Surprisingly, this is not the case as shown below.

To highlight the new physics involved in the multi-chromatic QP
lattices with second-neighbor hopping, we study the light excited in
photonic lattices with $\alpha=(\sqrt{5}-1)/2$. The unit of energy
is scaled by the nearest-neighbor hopping constant $c_1$.
Figure\,\ref{fig1} (b) shows the IPR values of all eigenstates as a
function of the potential strength $\lambda$.
The phase diagram in Fig.\,\ref{fig1} (b) provides three parts
separated by $\lambda_{c1}$ and $\lambda_{c2}$, $i.e.$, pure
delocalization phase, coexistence phase
\uppercase\expandafter{\romannumeral1}, and coexistence phase
\uppercase\expandafter{\romannumeral2}~\cite{Notes3}. When the
modulation strength $\lambda$ is sufficiently small
($<\lambda_{c1}$), IPR values are approximately vanishing for all
states, indicating that all states are extended. As $\lambda$ is
increased further, crossovers occur from purely delocalized to
coexistence phases (\uppercase\expandafter{\romannumeral1} and
further to \uppercase\expandafter{\romannumeral2}). It is remarkable
that the states are not simultaneously localized or extended, but
depend on their eigenenergies in contrast with the original AA
model.
The two coexistence phases \uppercase\expandafter{\romannumeral1}
and \uppercase\expandafter{\romannumeral2} are separated by the
critical value $\lambda_{c2}=2$ as seen from Fig.\,\ref{fig1}\,(b).
Note that in the coexistence phase
\uppercase\expandafter{\romannumeral1} two mobility edges come to
appear firstly (see the close-up view Fig.~\ref{fig1} (c)), followed
by a single mobility edge until $\lambda_{c2}$.

Figures \ref{fig1}\,(d-f) show the IPR values of states at various
$\lambda$'s. As observed from Fig.~\ref{fig1}\,(e), in the
coexistence phase \uppercase\expandafter{\romannumeral1} the
extended states in high energy band coexist with the localized
states in low energy band. There is a transition from delocalization
to localization with decreasing energy. Instead, in the coexistence
phase \uppercase\expandafter{\romannumeral2}, the system reverses
the transition following a sequence of
localization-to-delocalization alongside the lowering of energy
[Fig.~\ref{fig1} (d)]. In this regime, the extended states appear in
the high energy regime. Such a unique reversal between
\uppercase\expandafter{\romannumeral1} and
\uppercase\expandafter{\romannumeral2} arises from the self-duality
of our model, which means that the localization property at
$\lambda$ corresponds inversely to that at $4/\lambda$ (see the
Supplemental Material). It is worth noting that the reversal of
phases is absent from the QP models with mobility edges in
Refs.~\cite{SarmaPRL1988,SarmaPRL2010,SoukoulisPRL1982} because of
the broken self-duality.

\begin{figure}[t]
\includegraphics[width=0.9\linewidth]{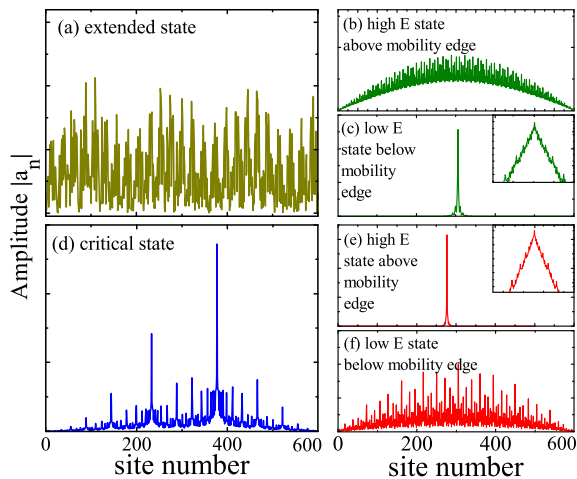}
\caption{(Color online) Typical eigenstates of our model in
different phases: (a) $\lambda=0.2$, (b, c) $\lambda=1.0$, (d)
$\lambda=2.0$, (e, f) $\lambda=3.0$, respectively. Insets are the
logarithmic plots of localized states, showing the exponentially
decaying tails $\log \left| a_n \right| \propto \exp(-\gamma x)$.
The parameters are the same as those in Fig.~\ref{fig1}
(b).}\label{fig2}
\end{figure}

Figures~\ref{fig2} summarizes the typical eigenstates at various
$\lambda$'s in order to confirm that the distinction of IPR does
guarantee various phases. In the purely delocalized regime, all
states extend over the entire system [Fig.~\ref{fig2}\,(a)]. In the
coexistence phase \uppercase\expandafter{\romannumeral1}, the states
are extended at higher energies [Fig.~\ref{fig2}\,(b)], while
exponentially localized at lower energies [Fig.~\ref{fig2}\,(c)],
which is the fingerprint of Anderson localization. However, in the
regime $\lambda>\lambda_{c2}$ the spatial behaviors of the states
are just the opposite of coexistence phase
\uppercase\expandafter{\romannumeral1} [Fig.~\ref{fig2} (e, f)].
At the critical strength $\lambda=2$, the states become spatially
fragmented [Fig.~\ref{fig2}\,(d)], which are intermediate between
spatially extended and exponentially localized. It should be
emphasized that this fragmented feature holds for any states at
$\lambda=2$.

\begin{figure}[htb]
\includegraphics[width=0.9\linewidth]{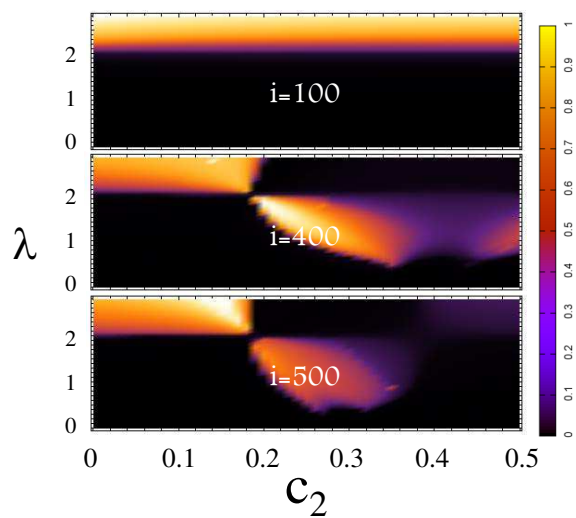}
\caption{(Color online) Dependence of IPR values of individual
eigenstates on $c_2$ and $\lambda$ for the lattices Eq.~(\ref{eq3}).
Shown are the eigenstates labeled by $i$. In the calculations the
lattices are kept being self-dual. Parameters: $N=600,
\alpha=(\sqrt{5}-1)/2$.}\label{fig3}
\end{figure}

To shed light on the dependence of localization properties on the
next-nearest-neighbor term, we plot in Fig.~\ref{fig3} the
distribution of IPR of several individual states on the
$c_2-\lambda$ plane. The developed intricate patterns illustrate the
nonuniform dependence of localization on $c_2$. For example, at a
fixed $\lambda$ smaller than $2$, when $c_2$ is increased the
$500$th state is transformed from delocalization to localization,
but further growth of $c_2$ brings it into the extended regime
again.

Diffusion of waves can provide the direct observables on the LD
transition and the mobility edges.
For light propagation in the photonic lattices with
next-nearest-neighbor hopping, equations of motion
read~\cite{LedererNature2003}
\begin{equation}
-i\frac{\partial a_n}{\partial z}=\beta_n a_n + c_1 (a_{n+1}+
a_{n-1})+ c_2 (a_{n+2}+ a_{n-2}),\label{eq5}
\end{equation}
where $z$ is the propagation coordinate of light, $\beta_n$ is the
single-site propagation constant of the underlying periodic lattice
defined by the multi-chromatic function. The equation is essentially
identical to the quantum description of non-interacting ultracold
atoms in the optical lattices, especially for shallow lattices, when
replacing the time variable $t$ by the propagation distance $z$.

\begin{figure}[htb]
\includegraphics[width=1.0\linewidth]{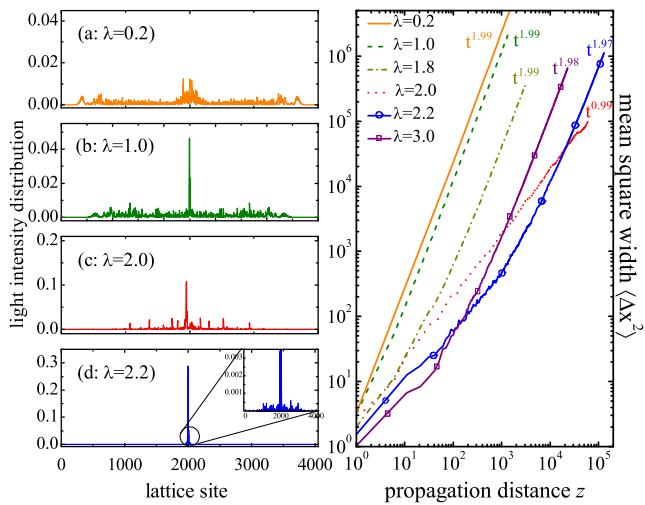}
\caption{(Color online) Left panels: Distributions of light
intensity after some time evolution in different localization
phases: (a) $\lambda=0.2$, (b) $\lambda=1.0$, (c) $\lambda=2.0$, and
(d) $\lambda=2.2$. The broadening of excitations in different
manners can be seen. Note the buildup of the central localized
components in (b, d). Right panel: Logarithmic plots of the averaged
mean square width $\langle \Delta x^2 \rangle$. The asymptotic
behavior of the spreading can be described by a power-law: $\langle
\Delta x^2 \rangle \sim t^{\gamma}$. The spreading exponent $\gamma$
provides a quantitative description of the quantum diffusion:
$\gamma=0$ corresponds to localization, $\gamma=1$ to normal
diffusion, and $\gamma= 2$ to ballistic motion.
}\label{fig4}
\end{figure}

The anomalous LD transition and the existence of mobility edge yield
a profound effect on the transport properties of the QP systems. We
inject a spatially narrow light beam into a single waveguide of the
lattice and monitor the time evolution to observe the signatures of
transition. Examples of light intensity during the evolution are
displayed in Fig.~\ref{fig4} (a-d). It is observed that the wave
functions grow in qualitatively different manners when $\lambda$ is
varied in different phases. In the regime of pure delocalization the
excitation is expanding continuously and the light intensity around
the input site is decaying gradually [Fig.~\ref{fig4}(a)], similar
to the intensity pattern in fully periodic arrays. Therefore, its
width increases ballistically, as illustrated by the solid line in
the right panel of Fig.~\ref{fig4}. In contrast, for the systems
belonging to the coexistence phase
\uppercase\expandafter{\romannumeral1} and
\uppercase\expandafter{\romannumeral2} [Fig.~\ref{fig4}(b, d)], a
twofold behavior of intensity distribution emerges. One sharp peak
localized around the initial position always persists during the
spreading, which indicates the existence of localized states.
Meanwhile, the ballistic tails contributed by the extended states
are superposed on the central peak. Because the fractions of
localized states in the spectrum are different, the heights of
central peaks are significantly different in phases
\uppercase\expandafter{\romannumeral1} and
\uppercase\expandafter{\romannumeral2}. For the case of critical
strength $\lambda=2$ the intensity structure becomes fragmented and
is composed of many separated spikes [Fig.~\ref{fig4}(c)].

Another quantity characterizing the transport of light is the mean
square width of wave packet $\Delta x^2 (z)=\sum\limits_i^N
(i-\langle i \rangle)^2 |a_i|^2$. The measure of localization
transition can be obtained by plotting $\left\langle \Delta x^2(z)
\right\rangle$, where $\langle ... \rangle$ denotes an average over
different excited positions. The right panel in Fig.~\ref{fig4}
shows how the mean square widths develop with the propagation
distance $z$ by varying $\lambda$ in different phases. In agreement
with the IPR picture obtained in Fig.~\ref{fig1}, in the pure
delocalization phase the wave packet exhibits a ballistic spreading
(solid line). For the coexistence phase
\uppercase\expandafter{\romannumeral1} and
\uppercase\expandafter{\romannumeral2}, on the other hand, the
widths never flatten off during the evolution even the existence of
localized modes, and the asymptotic behaviors also display the
ballistic motions. However, the light needs a longer transition
period from transient expansion to asymptotical stability, as
compared with the pure delocalization phase. Meanwhile, a new
transport behavior is observed when $\lambda=\lambda_{c2}$ (dot
line): the mean square width of light is proportional to the
propagation distance as the normal diffusion. Globally, increasing
the level of modulation leads to the transition from ballistic
motion to normal diffusion, and back to ballistic motion. This
clearly indicates there exists the LD transition qualitatively
different from that of the AA model~\cite{HiramotoJPSJ1988}.

Besides the asymptotic behavior of the spreading, we have analyzed
the transient process of light waves, especially in the coexistence
phases \uppercase\expandafter{\romannumeral1} and
\uppercase\expandafter{\romannumeral2}. In both regimes, the
quantity $\langle \Delta x^2 (z) \rangle$ approaches asymptotically
to be ballistic as $z \rightarrow \infty$. However, the two regimes
are contrasted in the transient behaviors about intersection point
depending on $\lambda>\lambda_{c2}$ or $\lambda<\lambda_{c2}$. In
the coexistence phase \uppercase\expandafter{\romannumeral1}, the
transient region is very narrow and the slopes of curves decline
gradually with the increase of modulation (dashed and dot-dashed
lines). On the other hand, we find a wide transient region when
$\lambda$ exceeds $\lambda_{c2}$. After a comparatively slow
evolution, the spreading curve of $\lambda=3.0$ surpasses that of
$\lambda=2.2$ (solid lines marked with $\square$ and {\footnotesize
$\bigcirc$)}. This leads to an intersection point of curves
belonging to the coexistence phase
\uppercase\expandafter{\romannumeral2}. This transient behavior is
quite abnormal~\cite{HiramotoJPSJ1988}, and can be used to
distinguish the coexistence phase
\uppercase\expandafter{\romannumeral1} and
\uppercase\expandafter{\romannumeral2}. Combining the intensity
distribution, asymptotic behavior and transient process, all being
able to be measured experimentally, constitutes a direct observation
of the anomalous transition in the QP photonic lattices.

Herein are some comments. We have also performed the calculations
for other incommensurate ratios, such as silver mean and bronze
mean, and observed qualitatively the same LD transition. But then we
should emphasize that in our QP system the nature of the LD
transition with the energy-dependent mobility edges is substantially
different from the usual 3D Anderson localization problem. In the 3D
disordered systems the phase transition is smooth through the
mobility edge $E_c$, so the states near $E_c$ are critical. However,
in our results the states near the mobility edges present a sudden
change from localization to delocalization, not being critical like
the 3D Anderson model.

The advances in photonic lattices allow for the realization of QP
potentials and the direct observation of light propagation as
well~\cite{Szameit}. In fact, the QP photonic lattices have been
fabricated for demonstrating the transition associated with the AA
model~\cite{SilberbergPRL2009}. The next-nearest-neighbor
interaction can be realized by constructing the zigzag structure of
arrays allowing the precise tuning of the
hopping~\cite{DreisowOL2008,WangOL2010}.
Besides light waves, ultracold atoms loaded into optical lattices
with high degree of control have provided another experimental
platform to observe the localization-related
phenomena~\cite{LSPNP2010,Modugno}, where the bichromatic optical
lattices were designed to implement the AA
model~\cite{RoatiNature2008}.
The QP potential in Eq.~(\ref{eq3}) can be produced by the
three-incommensurate-frequencies generalization of conventional
bichromatic lattices: superimposing three optical standing waves
with different wavelengths. Illuminating another two weaker laser
with the wavelengths being doubled into the main lattice, the beams
interfere to create a 1D multi-chromatic lattice with a certain
amount of incommensuration. The degree of incommensuration in the
main lattices can be adjusted by tuning the intensity of the
auxiliary lasers. To enter the regime beyond the nearest-neighbor
hopping, the optical potential formed by the main lattice should be
relatively shallow~\cite{HolthausPRA2007}.

In conclusion, we have investigated the self-dual multi-chromatic
quasiperiodic lattices with long-range hopping. Our model definitely
realizes the coexistence of extended and localized states, and the
LD transition becomes energy-dependent. We have demonstrated that
the self-duality of quasiperiodic lattice does not necessarily mean
pure spectrum of extended or localized states only. Instead,
extended and localized states can coexist at a fixed configuration
even if the self-duality persists.
AA model is just a special case of the self-dual quasiperiodic
models. As a further step, an interesting extension would be to
consider the nonlinear interactions, which may lead to anomalous
diffusion of quantum waves particularly in the coexistence
phase~\cite{ModugnoPRL2011}.


G.W. acknowledges fruitful discussions with Prof. Jiping Huang. N.L.
is supported by the National Natural Science Foundation of China
Grant No. 11205114. T.N. thanks the hospitality of Max Planck
Institute for the Physics of Complex Systems during the stay in
2012/13.

\end{document}